\documentclass[preprint,showpacs,preprintnumbers,amsmath,amssymb,showpacs,11pt,graphicx]{revtex4}
\usepackage{graphicx}% Include figure files
\usepackage{dcolumn}% Align table columns on decimal point
\usepackage{bm}% bold math
\usepackage{amssymb}

\newcommand{\be}{\begin{equation}}
\newcommand{\ee}{\end{equation}}
\newcommand{\ba}{\begin{eqnarray}}
\newcommand{\ea}{\end{eqnarray}}

\def\bea{\begin{eqnarray}}
\def\ena{\end{eqnarray}}

\renewcommand{\L}{{\cal L}}

\begin{document}

\title{Coincidence Problem in YM Field Dark Energy Model}

\author{Wen Zhao\footnote {wzhao7@mail.ustc.edu.cn} and Yang Zhang}
\affiliation{Astrophysics Center, University of Science and
Technology of China, Hefei, Anhui, China }

%\date{\today}

\begin{abstract}
{The coincidence problem is studied in the effective Yang-Mills
condensate dark energy model. As the effective YM Lagrangian is
completely determined by quantum field theory, there is no
adjustable parameter in this model except the energy scale, and
the cosmic evolution only depends on the initial conditions. For
generic initial conditions with the YM  condensate subdominant to
the radiation and matter, the model always has a tracking
solution, the Universe transits from matter-dominated into the
dark energy dominated stage only recently $z\sim 0. 3$, and evolve
to the present state with $\Omega_{y}\sim 0.73$ and $\Omega_m\sim
0.27$. }

\end{abstract}

%PACS numbers: 98.80.-k ~ 98.80.Es ~ 04.30.-w  ~04.62.+v

\pacs{98.80.-k, 98.80.Es, 04.30.-w, 04.62.+v }

%Key words: Yang-Mills field, dark energy

\maketitle

%\newpage
%\twocolumn
%\baselineskip=20truept

%\newpage

\section{Introduction}

Recent observations on the Type Ia Supernova (SNIa)\cite{sn},
Cosmic Microwave Background Radiation (CMB)\cite{map} and Large
Scale Structure (LSS)\cite{sdss} all suggest a flat Universe
consisting of dark energy (73\%), dark matter (23\%) and baryon
matter (4\%). How to understand the physics of the dark energy is
an important issue, having the equation of state  $\omega<-1/3$
and causing the recent accelerating expansion of the Universe. The
simplest model is the cosmological constant $\Lambda$ with
$\omega_{\Lambda}=-1$ (which can be viewed as the vacuum energy)
and fits the observation fairly well. However, throughout the
history of the Universe, the densities of matter and vacuum energy
evolve differently, so it appears that the conditions in the early
Universe have to be set very carefully in order for the energy
densities of them to be comparable today. This  is  the
coincidence problem \cite{coin}. Another one, called the
``fine-tuning" problem, is why the present vacuum energy density
is very tiny compared to typical scales in particle physics. These
problems have stimulated a number of approaches to build the dark
energy models with a dynamic field.

One class of approaches to the dark energy is to introduce dynamic
scalar fields with a tracker behavior. The scalar fields are
subdominant during early stages of expansion, and at late times,
they run into the attractor behaving as the dark energy, and
dominate over the matter component. The transition  to the state
$\omega \sim -1$ usually happens during the matter dominated era.
This kind of models include
 quintessence \cite{quint}  \cite{track_quint},
k-essence \cite{k}, phantom \cite{phantom}, quintom\cite{quintom}
etc. The quintessence fields have  attractor-like solutions
converging to a common, cosmic evolutionary track for a very wide
range of initial conditions rapidly. The initial value of
$\rho_{\phi}$ can vary by many orders of magnitude without
altering the cosmic history. The initial dark energy fraction
after inflation can be as large as $\Omega_{\phi}\sim  10^{-3}$,
which is natural in terms of the equipartition of energy among all
the dynamic degrees of freedom\cite{track_quint}. The k-essence
fields feature a tracker behavior during radiation dominated era,
and a $\Lambda$-like behavior shortly after the transition to
matter domination \cite{k}, but the initial dark energy fraction
only occupies a very narrow region in the phase space
\cite{track_k}.

The effective YM field condensate model has been introduced to
describe the dark energy \cite{z,Zhang,zhao}. It has interesting
features: the YM fields are the indispensable cornerstone to
particle physics,  gauge bosons have been observed. There is no
room for adjusting the form of effective YM Lagrangian as it is
predicted by quantum corrections according to field theory. In
this paper, we examine the  the coincidence problem in this model.
We find that the YM condensate field was subdominant during
earlier stages with a state of $\omega\sim 1/3$, and later it
turned into a state of $\omega \sim -1$ and only recently it
started to dominate over the matter. Moreover, the initial value
of YM energy fraction can be taken in a wide range of $\Omega_y
\simeq (10^{-20}, 10^{-2})$. Thus the model can solve the problem
naturally.

\section{The Effective Yang-Mills Field Model}

The effective YM  condensate cosmic model has been discussed in
Ref.\cite{z,Zhang,zhao}. The effective Lagrangian  up to 1-loop
order is \cite{pagels, adler}
 \be
 \L_{eff}=\frac{b}{2}F\ln\left|\frac{F}{e\kappa^2}\right|, \label{L}
 \ee
where  $b=11N/24\pi^2$ for the generic gauge group $SU(N)$ is the
Callan-Symanzik coefficient \cite{Pol}
$F=-(1/2)F^a_{\mu\nu}F^{a\mu\nu}$ plays the role of the order
parameter of the YM condensate,
 $e\simeq2.72$,
$\kappa$ is the renormalization scale with the dimension of
squared mass, the only model parameter. The attractive features of
this effective YM Lagrangian include the gauge invariance, the
Lorentz invariance, the correct trace anomaly, and the asymptotic
freedom\cite{pagels}. With the logarithmic dependence on the field
strength, $\L_{eff}$ has a form similar to he Coleman-Weinberg
scalar effective potential\cite{coleman}, and the Parker-Raval
effective gravity Lagrangian\cite{parker}. The effective YM
condensate was firstly put into the expanding Robertson-Walker
(R-W) spacetime to study inflationary expansion \cite{z} and the
dark energy \cite{Zhang}. We work in a spatially flat R-W
spacetime with a metric
 \be
 ds^2=a^2(\tau)(d\tau^2-\delta_{ij}dx^idx^j),\label{me}
 \ee
where $\tau=\int(a_0/a)dt$ is the conformal time. Assume that the
Universe is filled with the YM condensate. For simplicity we
study the $SU(2)$ group and consider the electric case with $B^2
\equiv0 $. The  energy density and pressure are given by \be
\label{rho}
 \rho_y=\frac{E^2}{2}\left(\epsilon+b\right),
 ~~~~p_y=\frac{E^2}{2}\left(\frac{\epsilon}{3}-b\right),
 \ee
where the dielectric constant is given by
 \be
 \epsilon=b\ln\left|\frac{F}{\kappa^2}\right|.\label{epsilon}
 \ee
and the equation of state (EoS) is
 \be
 \omega=\frac{p_y}{\rho_y}= \frac{\beta-3}{3\beta+3},\label{13}
 \ee
where $\beta\equiv\epsilon/b=\ln|\frac{E^2}{\kappa^2}|$. At the
critical point with the condensate order parameter $F=\kappa^2$,
one has $\beta=0$ and $\omega=-1$, the Universe is in  exact de
Sitter expansion \cite{z}. Around this critical point, $F<
\kappa^2$ gives $\beta<0$ and $\omega<-1$, and $F> \kappa^2$ gives
$\beta>0$  and  $\omega>-1$. So in the YM field model, EoS of
$\omega >-1$ and $\omega<-1$ all can be naturally realized. When
$\beta\gg1$, the YM field has a state of $\omega=1/3$, becoming a
radiation component. As is known, an effective theory is a simple
representation for an interacting  quantum system of many degrees
of freedom at and around its respective low energies. Commonly, it
applies only in low energies. However, it is interesting to note
that the YM condensate model as an effective theory intrinsically
incorporates the appropriate states for both high and low
temperature. As has been shown above, the same expression in
Eq.(\ref{rho}) simultaneously gives $p_y\rightarrow -\rho_y$  at
low energies, and $p_y\rightarrow \rho_y/3$  at high energies.
Therefore, our model of effective YM condensate can be used even
at higher energies than the renormalization scale $\kappa$.

The effective YM equations are
 \be
 \partial_{\mu}(a^4\epsilon~
 F^{a\mu\nu})+f^{abc}A_{\mu}^{b}(a^4\epsilon~F^{c\mu\nu})=0,
 \label{F1}
 \ee
the $\nu=0$ component of which is an identity, and the $\nu=1,2,3
$ spatial components of which reduce to
 \be
 \partial_{\tau}(a^2\epsilon E)=0.
 \ee
At the critical point ($\epsilon=0$), this equation is an
identity. When $\epsilon\neq 0$, this equation has an exact
solution as follows \cite{zhao},
 \be
 \beta~ e^{\beta/2}\propto a^{-2},\label{16}
 \ee
where the coefficient of proportionality in the above depends on
the initial condition. Near the critical point with
$|\omega+1|\ll1$, i.e. $\beta\ll1$, Eq.(\ref{16}) yields
 \be
 \beta\propto a^{-2},
 \ee
and the EoS is
 \be
 \omega+1\simeq\frac{4\beta}{3}\propto a^{-2}.\label{o+1}
 \ee
This  simple analysis shows that  with the expansion of the
Universe $\omega$ will goes to the critical point of $\omega=-1$,
an important character of the YM dark energy model. The YM
condensate can achieve the states of $\omega>-1$ and $\omega<-1$,
but it can not cross over $-1$, just like in the scalar models
\cite{-1}. At the early stages, $a \rightarrow 0$, Eq.(\ref{16})
leads to $\beta\gg1$, and Eq.(\ref{13}) gives
$\omega\rightarrow1/3$, so the YM condensate behaves as the
radiation component.

We should fix the value of $\kappa$, the only parameter in our
model. At the present time, the YM energy density
 \be
 \rho_y=\frac{bE^2}{2}(\beta+1)\simeq\frac{b\kappa^2}{2},
 \ee
and, as the dark energy, it should be $ \Omega_{y}\rho_0$, where
the present total energy density in the Universe $\rho_0\approx
8.099 \,h^2\times 10^{-11}eV^4$. We choose $\Omega_{y}=0.73$ as
has been observed,  yielding
 \be \label{kappa}
 \kappa=3.57 \, h\times 10^{-5} eV^2.
 \ee
This  energy scale is low compared to typical energy scales in
particle physics. So the ``fine-tuning" problem is also present in
this model.

To be more specific about how the YM condensate evolves in the
expanding Universe, we look at an early stage when the Big Bang
nucleosynthesis (BBN) processes occur around a redshift $z\sim
10^{10}$ with an energy scale $\sim 1$MeV. To see how the
evolution of $\rho_y$  depends the the initial condition, we
introduce the ratio of energies of the two components
 \be
 r_b\equiv\left.\frac{\rho_y}{\rho_r}\right|_{z=10^{10}},
 \ee
where  $\rho_r$ is the radiation energy density. We consider
$r_b<1$, i.e. the YM condensate is subdominant to the radiation
component initially. Of course, the YM condensate evolves
differently for different initial values of $r_b$. Nevertheless,
we will see that, as the result of evolution, the present Universe
is always dominated by the YM condensate $\Omega_y \sim 0.73$ for
a very wide range of initial values $r_b$.

 \begin{figure}
 \centerline{\includegraphics[width=10cm]{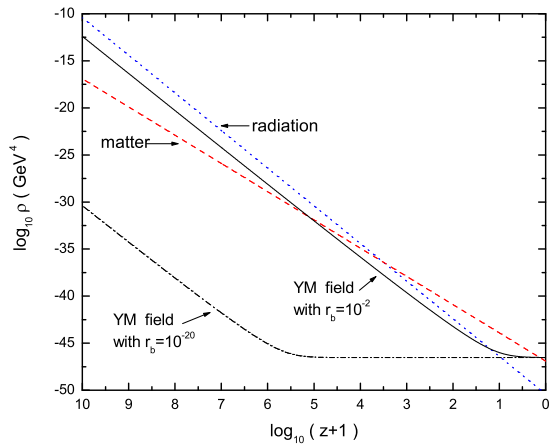}}
 \caption{\small The evolution of the energy densities versus redshift $z$.
 The dot line is for  the radiation density,
 and the dash line for the matter density.
 The solid line is for the YM energy density with $r_b=10^{-2}$,
 and the dash-dot one for the YM energy density with $r_b=10^{-20}$. }
 \end{figure}

 \begin{figure}
 \centerline{\includegraphics[width=10cm]{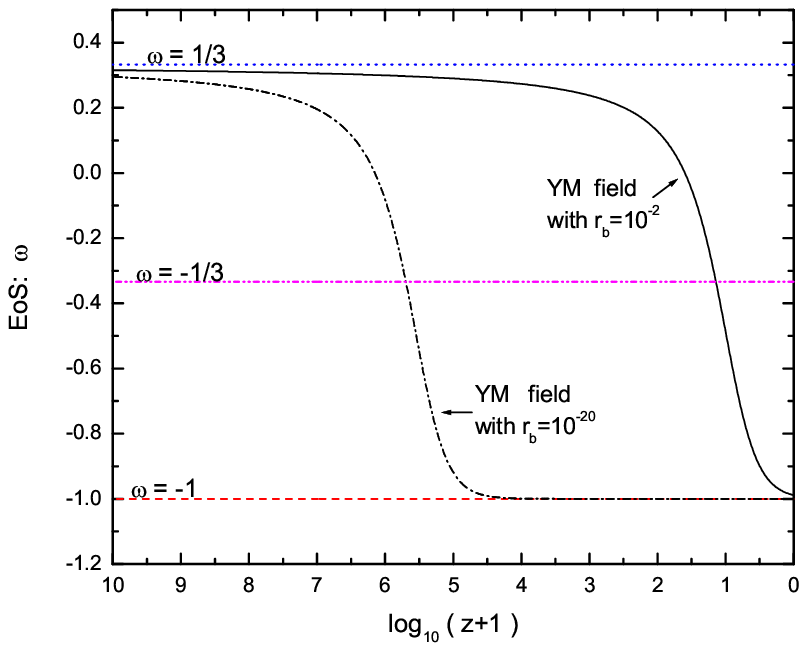}}
 \caption{\small The evolution of the equation of state $\omega$ for the YM field
 as a function of redshift $z$.
 Here the two curves are  for the same models as those in Fig. 1.}
 \end{figure}

Now we use the exact solution (\ref{16}) to plot the evolution of
$\rho_y$  as a function of the redshift $z$ in Fig.[1]. As
specific examples, here we take  $r_b=10^{-2}$, and
$r_b=10^{-20}$. In comparison, also plotted are the energy
densities of radiation, and of matter. It is seen that,  in the
early stages,  $\rho_y$  decreases as $\rho_y\propto a^{-4}$. So
the YM density is subdominant and tracks the radiation, a scaling
solution. The corresponding EoS of YM field is $\omega \simeq1/3$
shown in Fig.[2]. At late stages,  with the expansion of the
Universe, $a\rightarrow \infty$,
 $\beta$ decreases to nearly zero, and
$\omega\rightarrow-1$ asymptotically. Moreover,  this asymptotic
region is arrived at some redshift $z$ before the present time,
and this $z$ has different values for different initial values of
$r_b$. For smaller $r_b$, the transition redshift is larger (seen
in Fig.[2]), and the transition happens earlier. Once the
asymptotic region is achieved, the density of the YM field levels
off and remain a constant forever, like a cosmological constant.
We have also checked that the present value $\Omega_y \sim 0.73$
is also the outcome of the cosmic evolution for any value of $r_b$
in the very wide range $(10^{-20}, 10^{-2})$. So the coincidence
problem do not exist in the YM condensate model.

 \begin{figure}
 \centerline{\includegraphics[width=10cm]{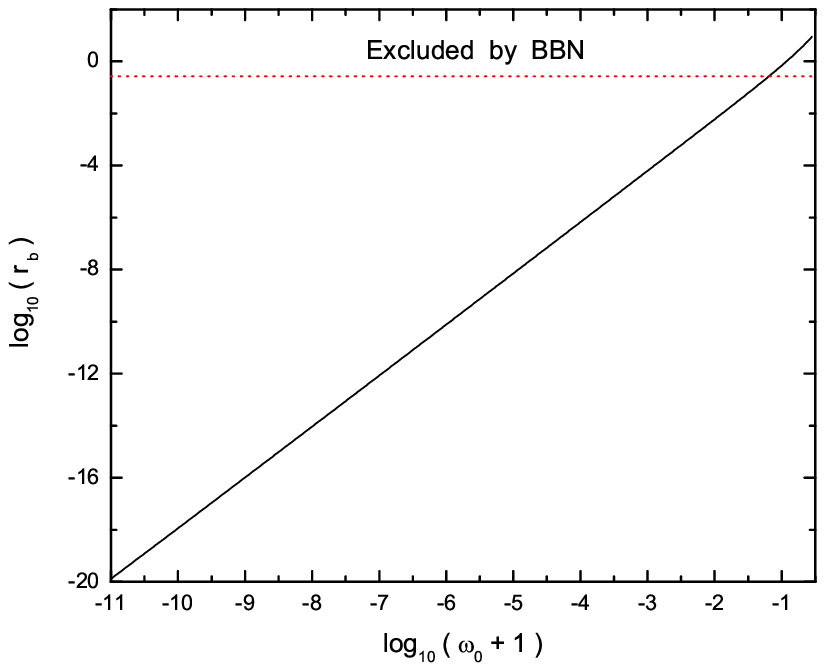}}
 \caption{\small The present EoS of the YM field $\omega_{0}$
 depends on the initial value $r_b$.
 The region above the dot line has been
 excluded by the observation of the BBN \cite{bbn}.}
 \end{figure}

 \begin{figure}
 \centerline{\includegraphics[width=10cm]{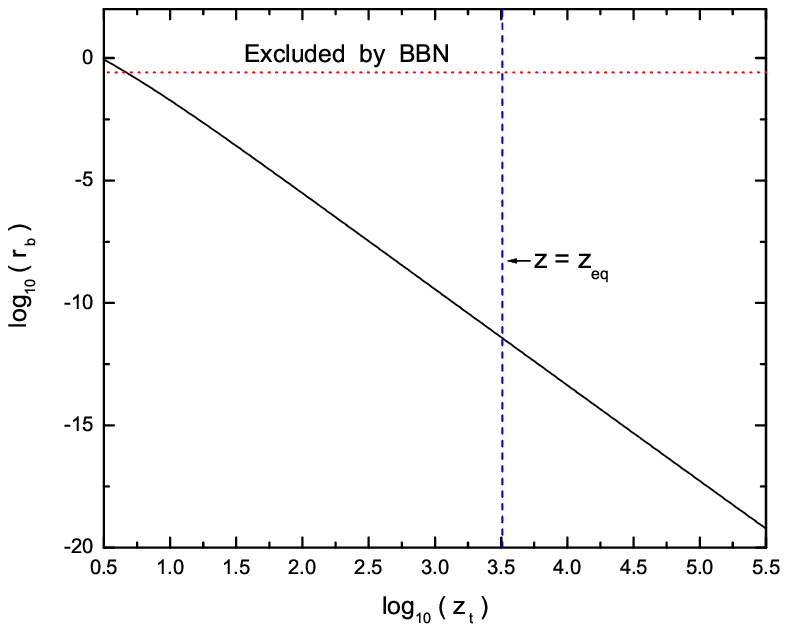}}
 \caption{\small The transition redshift $z_t$
 depends on the initial value  $r_b$.
 The region above the dot line has been
 excluded by the observation of the BBN \cite{bbn}. }
 \end{figure}

The present value of $\omega_0$  is nearly $-1$. Fig.[3] plots the
dependence of the present EoS $\omega_0$ on the initial condition
$r_b$. The  function $\log_{10}(r_b)$ versus
$\log_{10}(\omega_0+1)$ is nearly linear: a smaller $r_b$ leads to
a smaller $\omega_0$. For a value $r_b=10^{-2}$, one has
$\omega_{0}=-0.99$. For a value $r_b=10^{-20}$, $\omega_{0}$ would
be $-1$ accurately  up to one in $10^{11}$. Therefore, at present
the YM condensate  is very similar to the cosmological constant.

The solution in Eq.(\ref{16}) can converted into the following
form
 \be
 z=\sqrt{\frac{\beta}{\beta_0}}\exp\left[\frac{\beta-\beta_0}{4}\right]-1,
 \ee
where $\beta_0$ is the value of $\beta$ at $z=0$, depending on the
initial value $r_b$. For a fixed $\beta_0$, this formula tells a
one-one relation between the EoS (through $\beta$) and the
corresponding redshift $z$. As is seen from Fig.[2], the
transition of $\omega$ from $1/3$ to $-1$ occurs during a finite
period of time, instead of instantly. To characterize the time of
transition, we use $z_t$ to denote the redshift when
$\omega=-1/3$, i.e. $\beta=1$,  as given by Eq.(\ref{13}). This
is,  in fact, the time when the strong energy condition begins to
be violated, i.e., $\rho_y+3 p_y \leq 0$. Then
 \be
 z_t=\sqrt{\frac{1}{\beta_0}}\exp\left[\frac{1-\beta_0}{4}\right]-1.
 \ee
Therefore, this gives a function $z_t=z_t(r_b)$. Fig.[4] shows
how the transition redshift $z_t$ depends on the ratio $r_b$.
Interestingly, this transition can occur before, or after the
radiation-matter equality ($z_{eq}=3233$ \cite{map}). This feature
is different from the tracked quintessence or k-essence models in
which transition occurs during the matter dominated era
\cite{track_quint} \cite{k}. A larger $r_b$ leads to a smaller
$z_t$. For example, $r_b=10^{-2}$ leads to  $z_t\simeq 12.4\ll
z_{eq}$, and the transition occurs in the matter dominated stage,
and $r_b=10^{-20}$ leads to $z_t \simeq 5.0\times10^{5}\gg
z_{eq}$, and the transition occurs in the radiation dominated
stage.

The  value of  $r_b$ can not be chosen to arbitrarily large. In
fact, there is a constraint from the observation result of the
BBN. As is known, the presence of dark energy during
nucleosynthesis epoch will speedup the expansion, enhancing  the
effective species $N_{\nu}$ of neutrinos \cite{bbn_theory,bbn}.
 The latest analysis gives a constrain on
 the  extra neutrino species $\delta N_{\nu}\equiv N_{\nu}-3 <1.60$ \cite{bbn}.
Here in our model, the dark energy is played by the YM field. By a
similar analysis, the ratio $r_b$ is related to $\delta N_{\nu}$
through $  r_b= \frac{7\delta  N_{\nu}/4}{10.75} $. This leads to
an upper limit $r_b<0.26$, the present EoS $\omega_{0}<-0.94$ by
Fig.[3], and the transition redshift $z_t>5.8$ by Fig.[4]. The
range of initial conditions $r_b\in (10^{-20}, 10^{-2})$ that we
have taken satisfies this constraint.

\section{Summary}

The effective YM condensate has the advantageous characters: the
YM fields are indispensable to particle physics, there is no room
for adjusting the functional form of the Lagrangian as it is
predicted by quantum  field theory. As a model for the cosmic dark
energy, it has no free parameters  except the present cosmic
energy scale, and the cosmic evolution only depends on the initial
conditions. This study has shown that, for a generic value $r_b$
in a wide range of $(10^{-20}, 10^{-2})$, the present  YM dark
energy will be in $|\omega_0 +1 |<  0.01$, and the Universe has
$\Omega_{y}\sim 0.73$ and $\Omega_m\sim 0.27$ as the result of
cosmic evolution. Thus,  the YM model intrinsically has  the
tracking solution as in scalar field models. In this sense the
model can naturally solve  the coincidence problem.

It is found that at the early stages the subdominant YM field
behaves like a relativistic radiation with $\omega \simeq 1/3$ .
Later around a redshift $z_t$,
 the YM field transits into  the states $\omega \sim -1$,
and only recently it becomes dominant and drives  the current
accelerating expansion of the Universe. Interestingly, the YM
field model differs from the scalar models in that, the transition
of the state from $\omega\sim 1/3$ to $\omega \sim -1$ can occur
during radiation or  matter dominated era, depending  on the
initial condition $r_b$. Finally, the observation of BBN gives a
constraint $r_b<0.26$, which is not stringent. We have to say that
the ``fining-tuning" problem is not solved here as in most of
other current models.

~

ACKNOWLEDGMENT: W.Zhao's work has been partially supported by
Graduate Student Research Funding from USTC, and Y. Zhang's
research work has been supported by the Chinese NSF (10173008),
NKBRSF (G19990754), and by SRFDP.

%%%%%%%%%%%%%%%%%%%%%%%%%%%%%%%%%%%%%%%%%%%%%%%%%%%%%%%%%%%%%%%%%%%%%%%%%%%%%%%%%%%%%%%%%%%%%%%%%%%%%%%%%%

%%%%%%%%%%%%%%%%%%%%%%%%%%%%%%%%%%%%%%%%%%%%%%%%%%%%%%%%%%%%%%%%%%%%%%%%%%
%%%%%%%%%%%%%%%%%%%%%%%%% APPENDICES %%%%%%%%%%%%%%%%%%%%%%%%%%%%%%%%%%%%%
%%%%%%%%%%%%%%%%%%%%%%%%%%%%%%%%%%%%%%%%%%%%%%%%%%%%%%%%%%%%%%%%%%%%%%%%%%

\end{document}